\documentclass{emulateapj}

\usepackage{bm}
\usepackage{amsmath,amssymb}

\newcommand{\beq}{\begin{equation}}
\newcommand{\beqa}{\begin{eqnarray}}
\newcommand{\eeq}{\end{equation}}
\newcommand{\eeqa}{\end{eqnarray}}

\newcommand{\simgt}{\lower.5ex\hbox{$\; \buildrel > \over \sim \;$}}
\newcommand{\simlt}{\lower.5ex\hbox{$\; \buildrel < \over \sim \;$}}

\shorttitle{Weak lensing simulations}
\shortauthors{M.Sato et al.}

\begin{document}

\title{Simulations of Wide-Field Weak Lensing Surveys I:\\
 Basic Statistics and Non-Gaussian Effects }



\author{Masanori Sato\altaffilmark{1},
 Takashi Hamana\altaffilmark{2},
 Ryuichi Takahashi\altaffilmark{1},
 Masahiro Takada\altaffilmark{3},
 Naoki Yoshida\altaffilmark{3},\\
 Takahiko Matsubara\altaffilmark{1},
 Naoshi Sugiyama\altaffilmark{1,3}}

\affil{\altaffilmark{1} Department of Physics, Nagoya University,
 Nagoya 464-8602, Japan}
\affil{\altaffilmark{2} National Astronomical Observatory of Japan,
  Tokyo 181-8588, Japan}
\affil{\altaffilmark{3} Institute for the Physics and Mathematics of the
 Universe, University of Tokyo, Chiba 277-8582, Japan}

\email{masanori@a.phys.nagoya-u.ac.jp}

\begin{abstract}
We study the lensing convergence power spectrum and its covariance for a
standard $\Lambda$CDM cosmology.
We run 400 cosmological $N$-body simulations and use the outputs to
perform a total of 1000 independent ray-tracing simulations.
We compare the simulation results with analytic model predictions.
The semi-analytic model based on \cite{2003MNRAS.341.1311S} fitting formula
underestimates the convergence power by $\sim 30\%$ at arc-minute
 angular scales.
For the convergence power spectrum covariance, the halo model
reproduces the simulation results remarkably well over a wide range of
 angular scales and source redshifts.
The dominant contribution at small angular scales comes from the sample
 variance due to the number fluctuations of halos in a finite survey volume.
The signal-to-noise ratio for the convergence
power spectrum is degraded by the non-Gaussian covariances by up to a factor
5 for a weak lensing survey to $z_s\sim 1$.
The probability distribution of the convergence power spectrum
estimators, among the realizations, is well approximated by a
$\chi^{2}$-distribution with broadened variance given by the
 non-Gaussian covariance, but has a larger positive tail.
The skewness and kurtosis have non-negligible values especially for
a shallow survey.
We argue that a prior knowledge on the full distribution may be needed to
obtain an unbiased estimate on the ensemble averaged band power at each
angular scale from a finite volume survey. 
\end{abstract}

\section{Introduction}
Weak gravitational lensing provides a unique probe of the mass distribution 
in the Universe. By detecting the so-called ``cosmic shear'', one can
directly measure the power spectrum of mass fluctuations on cosmological 
scales of tens or even hundreds of mega-parsecs.
The significant detections of cosmic shear signals has been reported by
various groups, 
\citep[e.g.][]{2000MNRAS.318..625B,2000astro.ph..3338K,2000A&A...358...30V,2000Natur.405..143W,2003ApJ...597...98H},
and its ability to constrain cosmological parameters has been shown
\citep[e.g.][]{2006ApJ...644...71J,2006A&A...452...51S,2008A&A...479....9F}.

Weak lensing can also be a powerful probe of the nature of dark energy.
The growth rate of mass fluctuations can be measured by
``lensing tomography''
\citep[e.g.][]{1999ApJ...522L..21H,2002PhRvD..65f3001H,2004MNRAS.348..897T} 
which in turn provides tight constraints on the dark energy equation of 
state. A number of wide-field surveys are planned for this purpose,
such as 
Subaru Weak Lens Survey \citep[][]{2006SPIE.6269E...9M}, 
the Panoramic Survey Telescope \& Rapid Response System 
(Pan-STARRS\footnote{http://pan-starrs.ifa.hawaii.edu/public/}),
 the Dark Energy Survey (DES\footnote{http://www.darkenergysurvey.org/}),
 the 
Large Synoptic Survey Telescope (LSST\footnote{http://www.lsst.org/}),
 and the Joint Dark Energy Mission (JDEM\footnote{http://jdem.gsfc.nasa.gov/}).

However, to attain the full potential of future surveys, it is of great
importance to employ adequate statistical measures of weak lensing for
estimating cosmological parameters, properly taking into account
correlations of the
observables between different angular scales and redshifts, 
i.e. the covariances.
Since most useful cosmological information in weak lensing is inherent in
small angular scales that are affected by nonlinear clustering regime,
the covariance is by nature non-Gaussian. 
\cite{2009PhRvD..79b3520I} argued that the use of an inaccurate
covariance matrix may result in a systematically biased parameter estimate.
However, a modeling of the covariance matrix requires an accurate
modeling of nonlinear structure formation, which is not so easy. 

There are several approaches to 
modelling
the covariance matrix for 
a given
set of cosmological and survey parameters.
The most accurate way to obtain predictions for weak-lensing
 surveys is to perform ray-tracing simulations through large-volume,
high-resolution $N$-body
 simulations of cosmic structure formation 
\citep{2000ApJ...530..547J,
2001MNRAS.327..169H,
2003ApJ...592..699V,
2004APh....22...19W,
2008MNRAS.391..435F,
2009A&A...497..335T,
2009A&A...499...31H}.
Over a range of angular scales of interest, 
the non-Gaussian effects can be significant in weak 
lensing measurements
\citep{2000ApJ...537....1W,2007MNRAS.375L...6S,2009A&A...502..721E}.
However, rather expensive calculations are needed to create many
 independent realizations in this method.
A less expensive way would be to use semi-analytic models that are based
 on, for instance, the so-called ``halo model approach''
\citep{2001ApJ...554...56C,2009MNRAS.395.2065T}.

The purpose of this paper is to study the convergence power spectrum and
 its covariance using ray-tracing simulations for a 
concordance
 $\Lambda$CDM cosmology.
In order to obtain an accurate covariance matrix, we perform 1000
 independent ray-tracing simulations.
We compare our simulation results with the halo model predictions 
for the covariance
 matrix.
We also study the cumulative signal-to-noise ratio for measuring the
 convergence power spectrum taking into account the non-Gaussian errors.
Recently, \cite{2008ApJ...686L...1L} studied the angular power spectrum 
of the SDSS galaxy distribution and showed that the 
 signal-to-noise ratio integrated over a range of multipoles is two
 orders of magnitude smaller than the case of Gaussian fluctuations.
Motivated by their finding, we examine how the cumulative
signal-to-noise ratio for the lensing power spectrum measurement is
degraded by non-Gaussian errors using our large number of simulation
realizations. 
Furthermore, we will study how the convergence power spectrum estimates are
 distributed in different realizations: 
we will study
the probability distribution of convergence power spectrum
 and then compute the higher-order moments, skewness and kurtosis.

The structure of this paper is as follows. In Section~\ref{sec:sim} we describe
the details of $N$-body simulations and ray-tracing simulations. 
In Section~\ref{sec:ps}, after defining the lensing power spectrum,
we show the simulation results for power spectrum estimation and then
compare the results
with the analytical prediction. 
In  Section~\ref{sec:cov} we study the power spectrum covariance 
using
the simulations and the halo model, and also estimate the expected
signal-to-noise ratio for the power spectrum measurement taking into
account the non-Gaussian errors. In Section~\ref{sec:prob} we study the
probability distribution of power spectrum estimators in our
simulations. Section~\ref{sec:conc} is devoted to conclusion and
discussion.

\section{Numerical Methods}
\label{sec:sim}
\subsection{The Cosmological $N$-body Simulations}
\label{sim}
We use the parallel Tree-Particle Mesh code
 {\it Gadget-2} \citep{2005MNRAS.364.1105S} in its full Tree-PM mode.
We employ $256^3$ particles 
for each of two different simulation volumes, $240$ and $480h^{-1}$Mpc
on a side, which are used for ray-tracing simulations for redshift
ranges of $z=[0,1]$ and $[1,3]$, respectively (see
Fig.~\ref{ray_design}). 
The smaller volume simulation for lower redshifts is used in order to
 have higher mass and spatial resolutions because 
 nonlinear clustering is more evolving at lower redshifts. 
We generate the initial conditions following the standard Zel'dovich
 approximation. In this step we employed 
the linear matter transfer function computed from
 {\it CAMB} \citep{2000ApJ...538..473L}.
The initial redshift is set to 
 $z_{\rm init}=40$ and 
50 for the large- and small-box simulations, respectively.
We perform 200 realizations for each sets
and hence have a total of 400 
realizations. 

We adopt the concordance $\Lambda$CDM model with matter density
 $\Omega_{m}=0.238$, baryon density $\Omega_{b}=0.042$, dark
 energy density
 $\Omega_{\Lambda}=0.762$ with equation of state parameter
 $w=-1$, spectral index $n_s=0.958$, the variance of the density
 fluctuation in a sphere of radius 8 $h^{-1}$Mpc
 $\sigma_8=0.76$, and 
Hubble parameter $h=0.732$. 
These parameters are consistent with the WMAP 3-year
 results \citep{2007ApJS..170..377S}.

We have checked that our simulation result for the convergence power
spectrum agrees with the result using $512^3$ particles within 5 $\%$
 at $l\simlt 10^4$.
This is sufficient for our purpose, which is to study the power spectrum
and the covariance
down to arcminute scales.

\subsection{Ray-tracing Simulations}
\begin{figure}
\epsscale{1.0}
\plotone{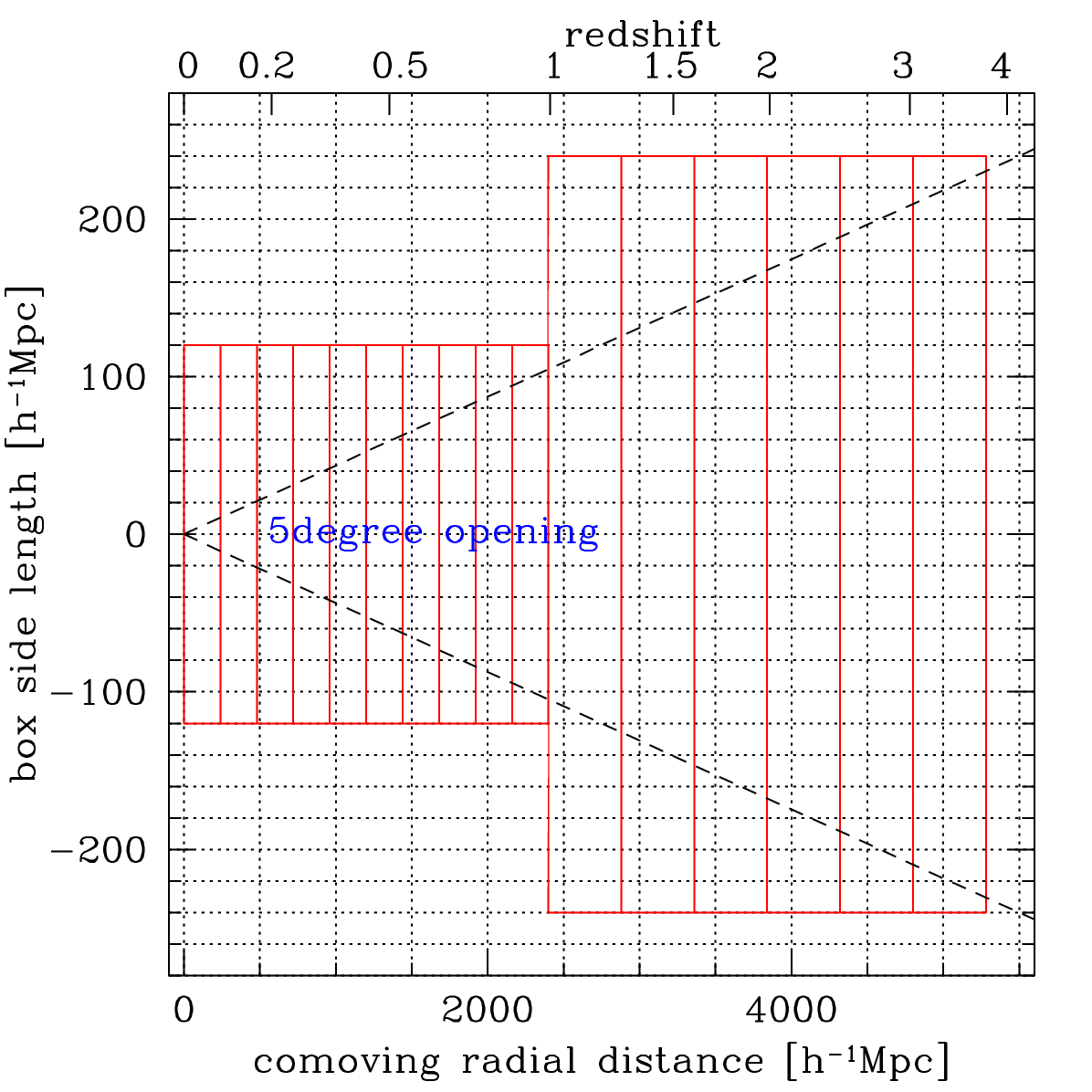}
\caption{The design of our ray-tracing simulations. Dashed lines show the
field-of-view 
spanning 
$\pm$2.5 degree.}
\label{ray_design}
\end{figure}

Fig.~\ref{ray_design} shows the design of our ray-tracing simulations.
We place the small- and large-volume simulations to cover a light cone 
of angular size $5^{\circ}\times 5^{\circ}$, from redshift 
$z=0$ to $z\sim 3.5$, 
using the tiling technique 
developed in 
\cite{2000ApJ...537....1W} and \cite{2001MNRAS.327..169H}.

We use the standard multiple lens plane algorithm in order to simulate the
distortion and magnification of background light rays by foreground matter.
Let us briefly describe the procedure to trace light rays through $N$-body
data \citep{2000ApJ...530..547J}.
In the standard multiple lens plane algorithm, the distance between
 observer and source is divided into $N$ intervals, separated by
 comoving distance $\Delta\chi$.
We adopt a fixed interval between lens planes by
 $\Delta\chi=120h^{-1}\rm Mpc$ (for this choice, the simulation box side
 lengths 
become multiples of $120h^{-1}\rm Mpc$).
The projected density contrast of the \textit{p}-th plane is given by
\begin{equation}
 \Sigma_p(\bm{\theta})
=\int_{\chi_{p-1}}^{\chi_{p-1}+\Delta\chi}\!
d\chi~\delta(\chi\bm{\theta},\chi) ,
\end{equation} 
where $\delta$ is the three-dimensional density fluctuation field along
the line-of-sight, 
$\delta\equiv \rho/\bar{\rho}-1$, and $\chi_{p-1}$ denotes the
 $\chi$-position of (\textit{p-1})-th lens plane, and
$\bm{\theta}$ is the two-dimensional vector denoting the angular position 
on the sky.
The projected density field is computed on $4096^2$ grids 
by projecting $N$-body particle distribution onto the lens plane 
based on 
the triangular-shaped cloud (TSC)
 assignment scheme \citep{1988csup.book.....H}.
An effective two-dimensional gravitational potential of the 
\textit{p}-th
 plane is related to projected density contrast via the two-dimensional
 Poisson equation  
\begin{equation}
 \nabla^2\psi_p=3\left(\frac{H_0}{c}\right)^2\Omega_m\Sigma_p.
\end{equation}
This equation is solved to compute $\psi_p$ using the
 fast Fourier transform method making use of the periodic boundary conditions.
Then, the first and second derivatives of $\psi_p$ are evaluated on 
each grid point.
Next, $2048^2$ rays are traced backward from the observer point.
The initial ray directions are set on $2048^2$ grids, which correspond
 to angular grid size of $5^{\circ}/2048\sim 0.15$ arcmin.
For each ray, we first computed ray positions on all the lens planes 
 using the lens equation:
\begin{equation}
\bm{\theta}_n=-\sum_{p=1}^{n-1}\frac{f_K(\chi_n-\chi_p)}{f_K(\chi_n)}
\nabla_{\bot}\psi_p+\bm{\theta}_1,
\end{equation}
where 
$f_K(\chi)$ is the comoving angular diameter distance given as function of
$\chi$; $f_K(\chi)=\chi$ for a flat universe.
The first and second derivatives of $\psi_p$ on a ray position are
 linearly interpolated from four nearest grids on which they were
 pre-computed.
The evolution equation of the Jacobian matrix, which describes
 deformation of an infinitesimal light ray bundle, is written as
\begin{equation}
\mathbf{A}_n=\mathbf{I}-\sum_{p=1}^{n-1}
\frac{f_K(\chi_p)f_K(\chi_n-\chi_p)}{f_K(\chi_n)}\mathbf{U}_p\mathbf{A}_p,
\label{Jacobian matrix} 
\end{equation}
where $\mathbf{I}$ is the identity matrix, and $\mathbf{U}_p$ is the shear
 tensor on the \textit{p}-th lens plane defined by
\begin{equation}
\mathbf{U}_{ij}\equiv\frac{\partial^2\psi_p}{\partial x_i\partial x_j},
\end{equation}
where $x_i\equiv \chi\theta_i$ and so on.
The Jacobian matrix is usually decomposed as 
\begin{equation}
 \mathbf{A}=\begin{pmatrix}
	     1-\kappa-\gamma_1 & -\gamma_2-w\\
	     -\gamma_2+w & 1-\kappa + \gamma_1
	     \end{pmatrix},
\end{equation}
where $\kappa$ is convergence, $|\gamma|=(\gamma^2_1+\gamma^2_2)^{1/2}$
 is the magnitude of the shear, and $w$ is a net beam rotation.  
Finally, the summation in Eq.~(\ref{Jacobian matrix}) yields
the Jacobian matrix.
The light-ray positions and four components of the Jacobian matrix on
 desired source planes are stored.
Source redshifts we consider in this paper are summarized
in Table.\ref{table1}

We perform 1000 ray-tracing realizations of the underlying density field
 by randomly shifting the simulation boxes assuming 
periodic boundary conditions.
In doing this, each simulation output is shifted in the same
way to make several lens planes to maintain the clustering pattern of
mass distribution within the simulation box\footnote{Note that, for
the separation between lens planes $\Delta\chi=120h^{-1}$Mpc, 
we make two (four) lens planes from one simulation output for the redshift
 range $z=[0,1]$ ($z=[1,3]$) because we use simulations of different
 volumes as described in \S~\ref{sim}.}
Importantly, however, in order to have {\it independent} realizations,
we do not use the same simulation output when making each ray-tracing
realization. 
Note that we use only one projection axis to all ray-tracing realizations,
unlike many previous works in which three orthogonal directions are used
to increase the number of {\it realizations}.

\begin{table}
\caption{
Source redshifts employed for our ray-tracing simulations. 
}
\label{table1}
\begin{center}

\begin{tabular}{cc}\hline  
$z_s$ & exact values \\ \hline
0.6 & 0.588542 \\
0.8 & 0.810822 \\
1.0 & 0.996884 \\
1.5 & 1.51902 \\
2.0 & 1.99765 \\
3.0 & 3.05725 \\ \hline
\end{tabular}
\end{center}
\end{table}

\section{Power Spectrum}
\label{sec:ps}

The mass density power spectrum $P_{\delta}(k)$ is defined as
\begin{equation}
 \langle\tilde{\delta}(\bm{k})\tilde{\delta^{*}}(\bm{k'})\rangle
  =(2\pi)^3\delta^3_D(\bm{k}-\bm{k'})P_{\delta}(k).
\label{3D_power}
\end{equation}
Likewise to Eq.~(\ref{3D_power}), one can define the convergence power spectrum $P_{\kappa}(l)$ as
\begin{equation}
 \langle\tilde{\kappa}(\bm{l})\tilde{\kappa^{*}}(\bm{l'})\rangle
  =(2\pi)^2\delta^2_D(\bm{l}-\bm{l'})P_{\kappa}(l).
\end{equation}
The conversion from the 3-D wave vector $\bm{k}$ to the 2-D angular wave
 vector $\bm{l}$ is done by the line of sight integration using
 the Limber approximation \citep{1954ApJ...119..655L,1998ApJ...498...26K}.
By using the Limber approximation, the convergence power spectrum is given
 by \citep[see, e.g.][]{2001PhR...340..291B,2003astro.ph..5089V}
\begin{equation}
 P_{\kappa}(l)=\int_0^{\chi_H}d\chi\frac{W(\chi)^2}{f_K(\chi)^2}P_{\delta}\left(\frac{l}{f_{K}(\chi)};\chi\right),
 \label{semi_pre}
\end{equation}
where $\chi_H$ is the horizon distance, defined as the comoving distance
 obtained for infinite redshift.
The weight function $W(\chi)$ is now
\begin{equation}
 W(\chi)=\frac{3}{2}\frac{H_0^2 \Omega_m}{c^2
  a(\chi)}f_K(\chi)\int_{\chi}^{\chi_H}d\chi' G(\chi')\frac{f_K(\chi'-\chi)}{f_K(\chi')},
\end{equation}
where $G(\chi)$ is the probability distribution of 
sources in
 comoving distance.
For simplicity, all sources are assumed to be located at the same
 redshift $z_s$, so that
\begin{equation}
 G(z)=\delta_D(z-z_s).
\end{equation}

\par

The binned convergence power spectrum 
can be estimated from each realization as
\begin{equation}
 \hat{P}_{\kappa}(l)=\frac{1}{N_l}\sum_{\bm{l};|\bm{l}|\in l
}|\tilde{\kappa}(\bm{l})|^2 ,
\label{eqn:ps_est}
\end{equation}
 where 
 the summation runs over modes whose lengths lie in the range
$l-\Delta l/2\le |\bm{l}_i|\le l+\Delta l/2$
for the assumed bin width $\Delta l$.
Throughout this paper we employ the bin width $\Delta \ln l=0.3$. 
The number of modes around a bin $l$ is approximately given by
\begin{equation}
N_l\approx A_{\rm s}\cdot \Omega_{\rm s}/(2\pi)^{2},
\end{equation}
 where $A_{\rm s}$ is the area of the two-dimensional shell 
around the bin $l$
and
can be given as $A_{\rm s}\approx 
2\pi l\Delta l + \pi(\Delta l)^2$, and $\Omega_{\rm s}$ is
the survey area. 
Taking the average of Eq.~(\ref{eqn:ps_est}) over
a number of realizations 
is expected to give the ensemble average expectation 
$P_{\kappa}(l)=\langle\hat{P}_{\kappa}(l)\rangle$.

\begin{figure}
\epsscale{1.0}
\plotone{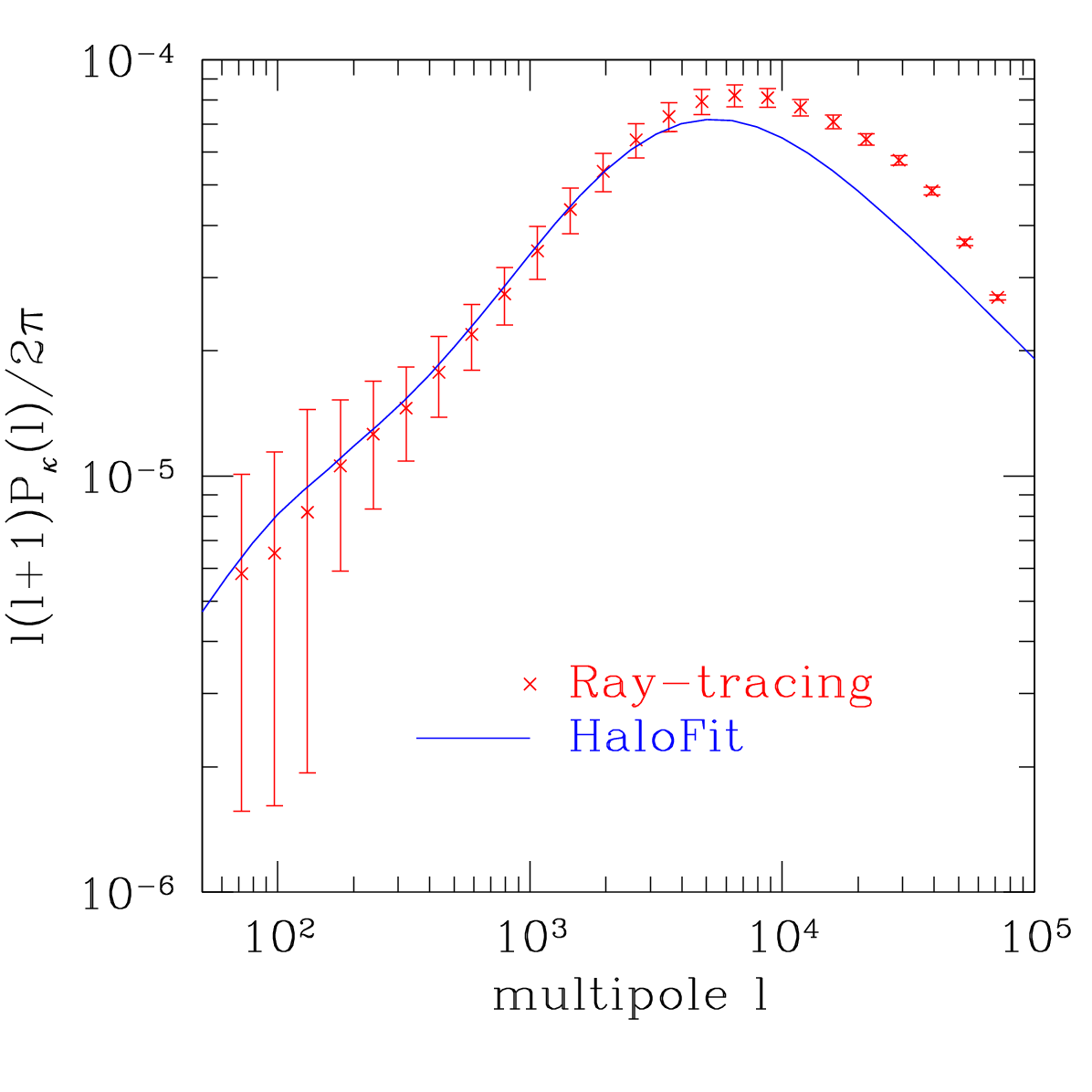}
\caption{Convergence power spectrum for sources at redshift $z_s=1.0$.
 The result from our 1000 ray-tracing simulations is shown as 
the cross symbols with 
 error bars ($\pm 1\sigma$ variance). 
We also show the semi-analytic prediction computed 
from  Eq.~(\ref{semi_pre}) 
using {\it HaloFit} to model the 3D mass power spectrum.
}
\label{power}
\end{figure}

Fig.~\ref{power} shows the convergence power spectrum obtained
from our ray-tracing simulations for $z_s=1.0$.
We compare it with the 
semi-analytic prediction 
computed using 
the \cite{2003MNRAS.341.1311S} fitting formula (hereafter {\it
HaloFit}) to compute the nonlinear 
matter power spectrum $P_{\delta} (k)$.
The {\it HaloFit} underpredicts the convergence power at intermediate 
and small scales, $l > 3000$.
A similar disagreement is also found 
in \cite{2009A&A...499...31H}
using Millennium Simulation \citep[][]{2005Natur.435..629S}
which has higher mass and spatial resolutions than ours. 

\begin{figure}
\epsscale{1.0}
\plotone{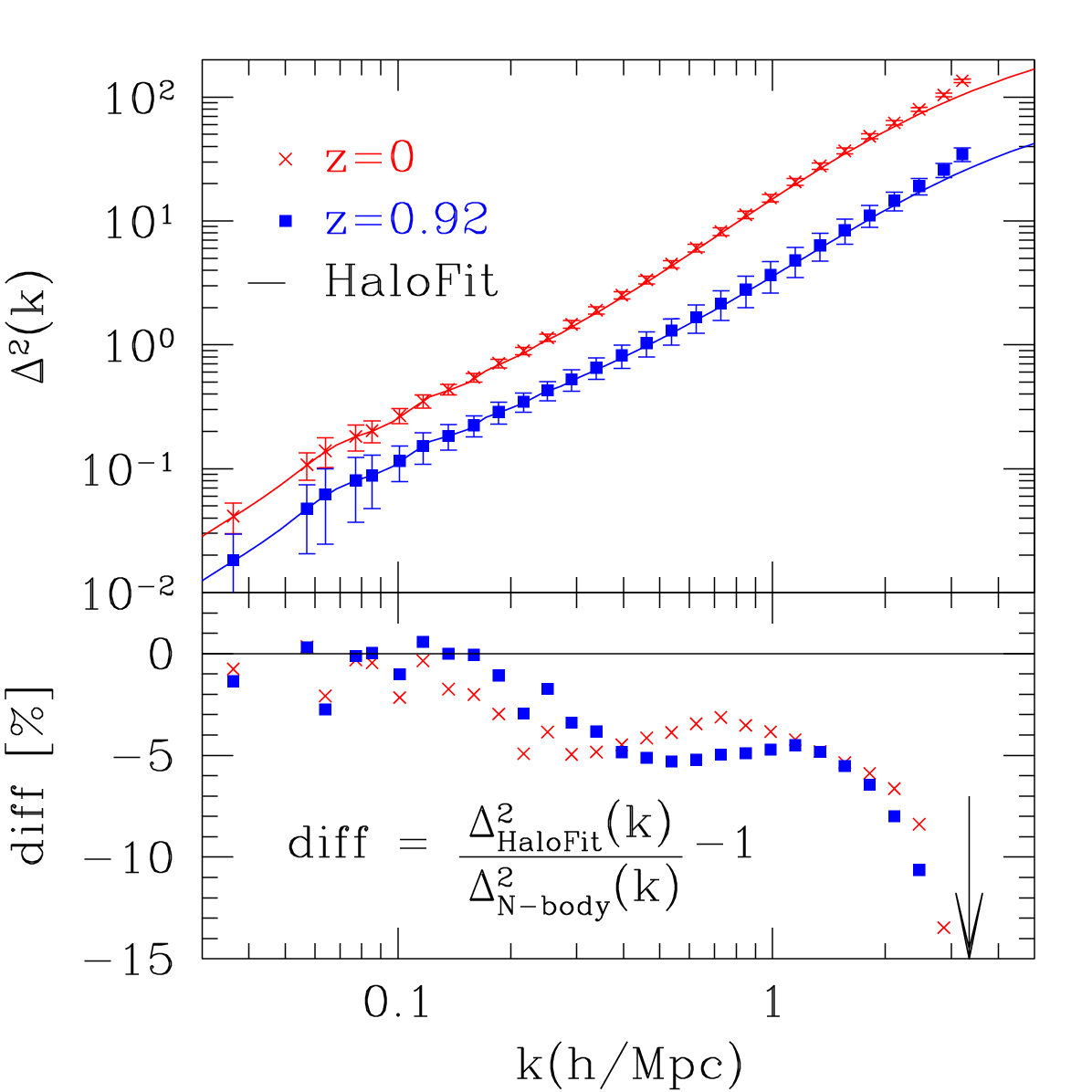}
\caption{ 
{\it Top panel} : Comparison of the dimensionless 3D mass 
power spectrum, $\Delta^2(k)=k^3P(k)/2\pi^2$, 
 obtained from 200 realizations to the {\it HaloFit} prediction at
 $z=0$ and $0.92$. {\it Bottom panel} : The fractional 
difference is shown in
 percent. The arrow shows the Nyquist wavenumber.}
\label{pk}
\end{figure}

To further explore the cause of this discrepancy, 
we 
study
the three-dimensional mass power spectra using our $N$-body simulations.
Fig.~\ref{pk} compares the 
power spectra obtained
from 200 realizations with the {\it HaloFit} predictions at
 $z=0$ and $0.92$.
The arrow shows the Nyquist wavenumber.
The {\it HaloFit} results are approximately $5-10\%$ lower than the results
 from simulations.  
\cite{2008arXiv0812.1052H} also report a similar disagreement.
It appears that the discrepancy in the convergence power at high $l$
is owing to inaccuracy in {\it HaloFit}.

\begin{figure}
\epsscale{1.0}
\plotone{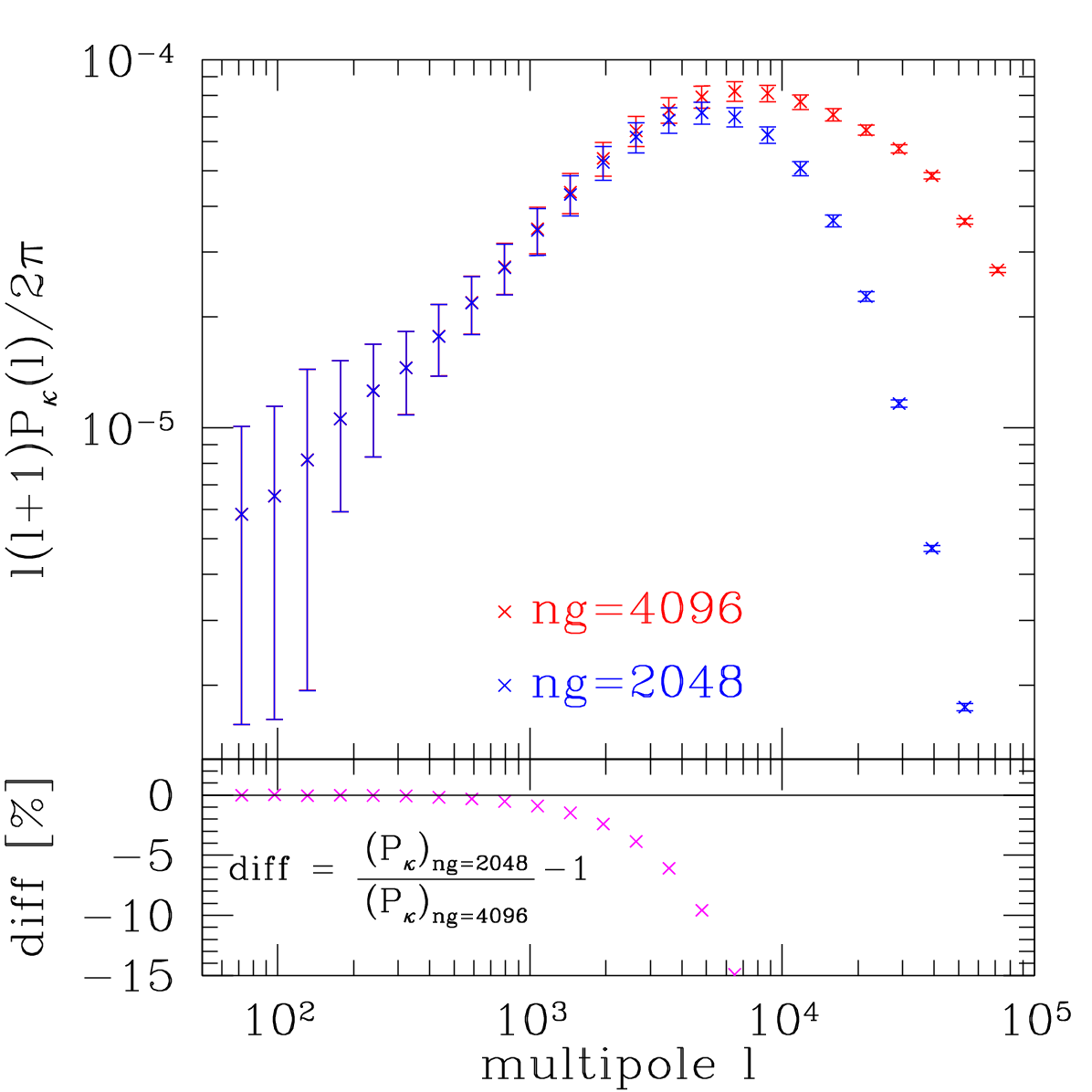}
\caption{
The effect of grid size in ray-tracing simulation on the power spectrum 
estimation. 
{\it Top panel} : The two spectra computed from 1000 realizations
 employing $2048^2$ and $4096^2$ grids. 
 {\it Bottom panel} : The fractional difference is shown in percent.
}
\label{smoothing}
\end{figure}

We also examine the effect of 
smoothing used in ray-tracing simulations
by varying 
the grid size. 
Fig.~\ref{smoothing} compares the convergence power spectra calculated
for two different 
grid sizes, 
$2048^2$ and $4096^2$ grids, respectively. 
Clearly, the coarser grid size yields a smaller power at $l > 3000$. 
In the bottom panel, we show that the difference reaches $\sim 5\%$ at $l\sim 3000$.
When we compute the projected density field, we would naively expect
that a finer grid size
 provides a higher resolution 
in the lensing convergence map, if the original $N$-body simulation has
a sufficient spatial resolution.
Further halving the grid size, i.e. changing the grid number to 
$8192^2$ from $4096^2$, is similarly thought to give a better accuracy
up to higher multipoles.
We have checked the difference reaches $\sim 5\%$ at $l\sim 6000$. 
Therefore, we conclude that 
estimating power spectrum 
from the  projected density fields 
on $4096^2$ grids is sufficiently accurate 
up to $l\sim 6000$.
In the following section, we focus on the
 power spectrum information up to $l\sim
 6000$.

\section{Covariance Matrix}
\label{sec:cov}
The covariance matrix of the convergence power spectrum between
 $P_{\kappa}(l)$ and $P_{\kappa}(l')$ is 
formally expressed as a sum of the Gaussian and
 non-Gaussian contributions
 \citep[e.g.][]{1999ApJ...527....1S,2001ApJ...554...56C}:
\begin{align}
 {\rm Cov}[P_{\kappa}(l),P_{\kappa}(l')]&\equiv\left\langle
  \left(\hat{P}_{\kappa}(l)-P_{\kappa}(l)\right)\left(\hat{P}_{\kappa}(l')-P_{\kappa}(l')\right)\right\rangle \nonumber\\
&=\frac{2}{N_{l}}P_{\kappa}(l)^2\delta^K_{l,l'}\nonumber\\
&+\frac{1}{\Omega_{\rm s}}\int_{l}\frac{d^2
 \bm{l}}{A_{\rm s}}\int_{l'}\frac{d^2\bm{l'}}{A_{\rm s'}}T(\bm{l},-\bm{l},\bm{l}',-\bm{l}'),
\label{covariance}
\end{align}
where $\delta^K_{l,l'}$ is the Kronecker delta function 
and $T$ is the lensing trispectrum defined as
\begin{equation}
 \langle\tilde{\kappa}(\bm{l}_1)\tilde{\kappa}(\bm{l}_2)\tilde{\kappa}(\bm{l}_3)\tilde{\kappa}(\bm{l}_4)\rangle 
\equiv (2\pi)^2\delta_D(\bm{l}_{1234})T(\bm{l}_1,\bm{l}_2,\bm{l}_3,\bm{l}_4),
\end{equation}
where we have introduced notation $\bm{l}_{1234}=\bm{l}_1+\bm{l}_2+\bm{l}_3+\bm{l}_4$.
In the Limber approximation, $T$ is a simple projection of the 
three-dimensional mass
trispectrum $T_{\delta}$ given by
\begin{align}
&T(\bm{l}_1,\bm{l}_2,\bm{l}_3,\bm{l}_4)=\int_0^{\chi_H}d\chi\frac{W(\chi)^4}{f_K(\chi)^{6}} \nonumber\\
&\times T_{\delta}\left(\frac{\bm{l}_1}{f_K(\chi)},\frac{\bm{l}_2}{f_K(\chi)},\frac{\bm{l}_3}{f_K(\chi)},\frac{\bm{l}_4}{f_K(\chi)};\chi\right).
\end{align}

In Eq.~(\ref{covariance}),
the first term describes 
the Gaussian error contribution that has vanishing correlations between 
different multipole bins,  
whereas the second term describes the non-Gaussian
 contribution arising from mode coupling due to nonlinear clustering. 
Both the terms scale with the survey area as $\propto 1/\Omega_{\rm s}$.
It should be also noted that the Gaussian term depends on the bin width $\Delta
 l$, whereas the non-Gaussian term does not (because
 $\int_{l}d^2\bm{l}/A_{\rm s}\approx 1$). Thus decreasing
 $\Delta l$ increases the Gaussian contribution 
relative to the non-Gaussian errors. 

\begin{figure}
\epsscale{1.15} \plotone{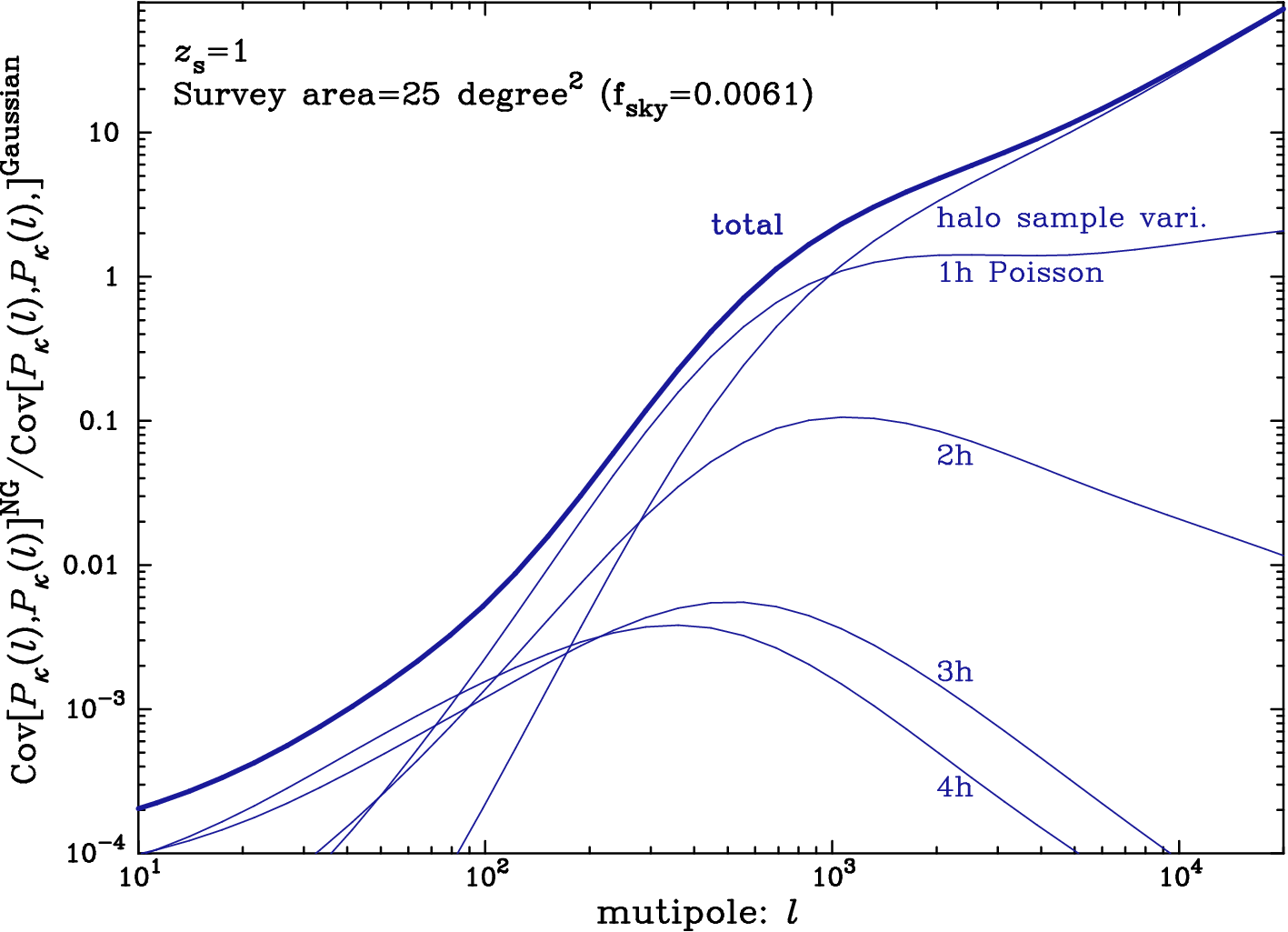} \caption{The halo model
predictions for the ratio of the diagonal non-Gaussian covariance
components to the Gaussian errors as a function of multipole, where the
Gaussian errors are  given by ${\rm Cov}^{\rm
G}=P_\kappa(l)^2/(f_{\rm sky}l^2\Delta\ln l )$. Note that we assume
source redshift $z_s=1$, the survey area $\Omega_{\rm s}=25$ sq. degrees
and the multipole bin width $\Delta\ln l=0.3$.  In our halo model the
non-Gaussian errors are given by the sum of the different halo terms of
trispectrum contribution (see Eqs.~\ref{covariance} and
\ref{eqn:halo_trisp}) and the halo sample variance (see
Eq.~\ref{eqn:1hsv}). The thick solid curve shows the total contribution
to the non-Gaussian errors, while the thin solid curves show each
different term contributions as indicated by each label.  }
\label{fig:halo_trisp}
\end{figure}
\subsection{Halo Model Approach for the Covariance}
To make an analytic estimation of 
the lensing power covariance using Eq.~(\ref{covariance}),
 we need to 
model the mass trispectrum that can 
account for the non-linear clustering 
at small angular scales.
In this paper we employ the halo model approach \citep[][also see 
\cite{2002PhR...372....1C} for a thorough
 review]{2000MNRAS.318..203S,2000ApJ...531L..87M,2000MNRAS.318.1144P}.
In the halo model, the power spectrum is given by a sum of two terms:
 the so-called 1-halo term which describes
correlation of dark matter
 particles within each halo, 
and the 2-halo term which describes correlation of
 particles in different halos.
Likewise, the trispectrum consists of four terms, from one to four halo
 terms \citep{2001ApJ...554...56C}:
\begin{equation}
 T_{\delta}=T_{\delta}^{1h}+T_{\delta}^{2h}+T_{\delta}^{3h}+T_{\delta}^{4h},
\label{eqn:halo_trisp}
\end{equation} 
where 
we have suppressed the arguments $(\bm{k}_1,\bm{k}_2,\bm{k}_3,\bm{k}_4)$
of $T_\delta$. 
These four terms contribute to the power at each $l$ differently.
The 1-halo term gives dominant contribution in the nonlinear clustering regime.

To complete the halo model approach, we need suitable models for the
three ingredients: the halo mass density profile, the mass function of
halos, and the biasing of halo distribution, each of which is specified by
halo mass $m$ and redshift $z$ for a given cosmological model. For these
we employ an NFW profile \citep{1997ApJ...490..493N}, and  the
fitting formulae for mass function and halo bias derived in
\cite{1999MNRAS.308..119S}. The details of our halo model implementation
can be found in \cite{2003MNRAS.340..580T} 
\citep[also see][]{2003MNRAS.344..857T}. 

However it turns out there is additional contribution to the
non-Gaussian covariance, which becomes significant on small scales as
described below.  
As first pointed
out in \cite{2003ApJ...584..702H}, 
the halo clustering causes 
additional sampling variance to the covariance due to the
statistical fluctuations in the number of halos sampled by a {\em finite}
survey volume \citep[also
see][]{2007NJPh....9..446T,
2006MNRAS.370L..66N,
2007ApJ...671...14Z,2009ApJ...698..143R}.  
In simpler words, if
massive halos happen to be more (less) in a surveyed region, the measured power
spectra would very likely have greater (smaller) amplitudes than
expected from the ensemble average.

According to the formulation developed in Appendix in
\cite{2007NJPh....9..446T} \citep[also see the discussion around Eq.~(7)
in][]{2007ApJ...671...14Z}, the sample variance to which we hereafter refer
as the halo sample variance (HSV)
is expressed as
\begin{eqnarray}
{\rm Cov}_{{\rm HSV}}[P_\kappa(l),P_\kappa(l')]&=&\int_0^{\chi_s}
\!\!d\chi\left(\frac{d^2V}{d\chi d\Omega}\right)^2\nonumber\\
&&\hspace{-5em}
\times\left[
\int\!\!dM\frac{dn}{dM}b(M)|\tilde{\kappa}_M(l)|^2
\right]
\nonumber\\
&&\hspace{-5em}
\times\left[
\int\!\!dM'\frac{dn}{dM'}b(M')|\tilde{\kappa}_{M'}(l')|^2
\right]
\nonumber\\
&&\hspace{-5em}
\times\int_0^{\infty}\!\!\frac{kdk}{2\pi}\!P_\delta^{\rm
L}(k;\chi)\left|
      \tilde{W}\!\left(k\chi\Theta_{\rm s }\right)
\right|^2,
\label{eqn:1hsv}
\end{eqnarray}
where $d^2V/d\chi d\Omega$ is the comoving volume per unit solid angle
and unit radial comoving distance, given as $d^2V/d\chi d\Omega=\chi^2$
for a flat universe, $\chi_s$ is the comoving distance to a source
redshift $z_s$, $dn/dM$ is the halo mass function, $b(M)$ is the halo
bias parameter, and $\tilde{\kappa}_M(l)$ is the angular
Fourier-transform of the convergence field for a halo with mass $M$
\citep[see Eqs.~28 and 31 in][for the definition]{2003MNRAS.340..580T}.
The quantity $P_\delta(k)$ is the linear 3D mass power spectrum, and
$\tilde{W}(x)$ is the Fourier transform of the survey window function;
for this we simply employ $\tilde{W}(k\chi\Theta_{\rm
s})=2J_1(k\chi\Theta_{\rm s})/(k\chi\Theta_{\rm s})$ ($J_1(x)$ is the
first-order Bessel function) assuming $\Theta_{\rm s}=\sqrt{\Omega_{\rm
s}/\pi}$ for a given survey area $\Omega_{\rm s}$. Thus we have assumed
that the survey area is sufficiently large and the number fluctuations
of halos are in the linear regime.  Note that this sampling variance
contribution does not necessarily scale with $1/f_{\rm sky}$ unlike
other covariance terms. The sample variance depends on $f_{\rm sky}$ via
the shape of linear power spectrum. For a CDM spectrum it becomes
smaller with increasing $\Omega_{\rm s}$, and decreases faster than the
other covariance terms that have the scaling of $\Omega_{\rm s}^{-1}$ or
$f_{\rm sky}^{-1}$ if $\Omega_{\rm s}$ is greater than a few hundreds
square degrees, for multipoles of interest ($l\simgt 1000$).

More exactly Eq.~(\ref{eqn:1hsv}) was derived by replacing the function
$S_{(b)}$ in Eq.~(B1) of \cite{2007NJPh....9..446T} with
$|\tilde{\kappa}_M(l;z)|^2$, which is the lensing power spectrum
contribution due to a halo with mass $M$ and at redshift $z$. In fact
the first term on the r.h.s. of Eq.~(B1) corresponds to the 1-halo term
of non-Gaussian errors in Eq.~(\ref{covariance}), while the second term
in Eq.~(B1) yields Eq.~(\ref{eqn:1hsv}). The contribution of
Eq.~(\ref{eqn:1hsv}) arises for any finite-volume survey because the
halo distribution has modulations due to the biased density fluctuations
over the survey window.  The full derivation of non-Gaussian covariance
within the context of the halo model approach is beyond the scope of
this paper and will be presented elsewhere.

To obtain a more physical insight, it would be useful to note that the
sample variance (\ref{eqn:1hsv}) is roughly expressed as
\begin{equation}
{\rm Cov}_{{\rm HSV}}\sim \bar{b}^2\sigma_{\rm rms}^2(\Theta_{\rm s})
P_\kappa^{\rm 1h}(l)P_\kappa^{\rm 1h}(l'), 
\label{eqn:1hsv_asymp}
\end{equation}
where $\bar{b}$ is the halo bias averaged over halo masses and redshift
interval, $\sigma_{\rm rms}(\Theta_{\rm s})$ is the rms of angular mass
density fluctuations for the survey area, and $P_\kappa^{\rm 1h}(l)$ is
the 1-halo term of the convergence power spectrum. Here the combination
of $\bar{b}\sigma_{\rm rms}(\Theta_{\rm s})$ gives the rms fluctuations
in the number of halos in the survey area. Thus the sample variance
strength is proportional to combined effect of the convergence power
spectrum and the number fluctuations of massive halos due to the
large-scale mass density fluctuations.  
The sample variance (\ref{eqn:1hsv}) is vanishing if the
halo distribution does not have any clustering, i.e. is completely
random (corresponding to the limit $\bar{b}\sigma_{\rm rms}\rightarrow 0$).
Note that, on the other hand,
the 1-halo term of the trispectrum accounts for the Poisson contribution
to the sample variance arising from the discreteness nature of halo
distribution.

Fig.~\ref{fig:halo_trisp} shows the halo model predictions for the
diagonal non-Gaussian covariances relative to the Gaussian errors as a
function of multipoles. Note that we consider source redshift $z_s=1$ as
a representative example, but the results are very similar for other
source redshifts we consider in this paper. The different halo term
contributions to the covariance are more important on small multipoles,
while the 1-halo term becomes increasingly significant with increasing
multipoles. Importantly the sample variance contribution due to the
number fluctuations of massive halos, given by Eq.~(\ref{eqn:1hsv}),
becomes dominant over other non-Gaussian errors at high multipoles
$l\simgt 1000$, boosting the non-Gaussian error strengths by an order of
magnitude up to $l\sim 10^4$ from the estimate without this new effect.
For high multipole limit, the ratio
of the diagonal non-Gaussian covariance to the Gaussian errors has an
asymptotic scaling as ${\rm Cov}^{\rm NG}/{\rm Cov}^{\rm
G}\sim \bar{b}^2\sigma^2_{\rm rms}P^{\rm 1h}_\kappa(l)P^{\rm 1h}_\kappa
/[2P^{\rm 1h}_\kappa(l)^2/N_l]\sim \bar{b}^2\sigma^2_{\rm
rms}(\Theta_{\rm s})N_l/2 \propto l^2\Delta\ln l$. Hence, ${\rm
Cov}^{\rm NG}/{\rm Cov}^{\rm G}\propto l^2$ for a constant bin width
$\Delta\ln l$ as implied from the results 
around $l\sim 10^4$ in Fig.~\ref{fig:halo_trisp}, because
$\bar{b}\sigma_{\rm rms}(\Theta_{\rm s})$ is constant for fixed survey
area and cosmological model.

This contribution has been ignored in previous studies 
 and we will
below study the impact on the lensing power spectrum covariance
 comparing the halo model predictions with the simulation
results. 

\begin{figure}
\epsscale{1.0}
\plotone{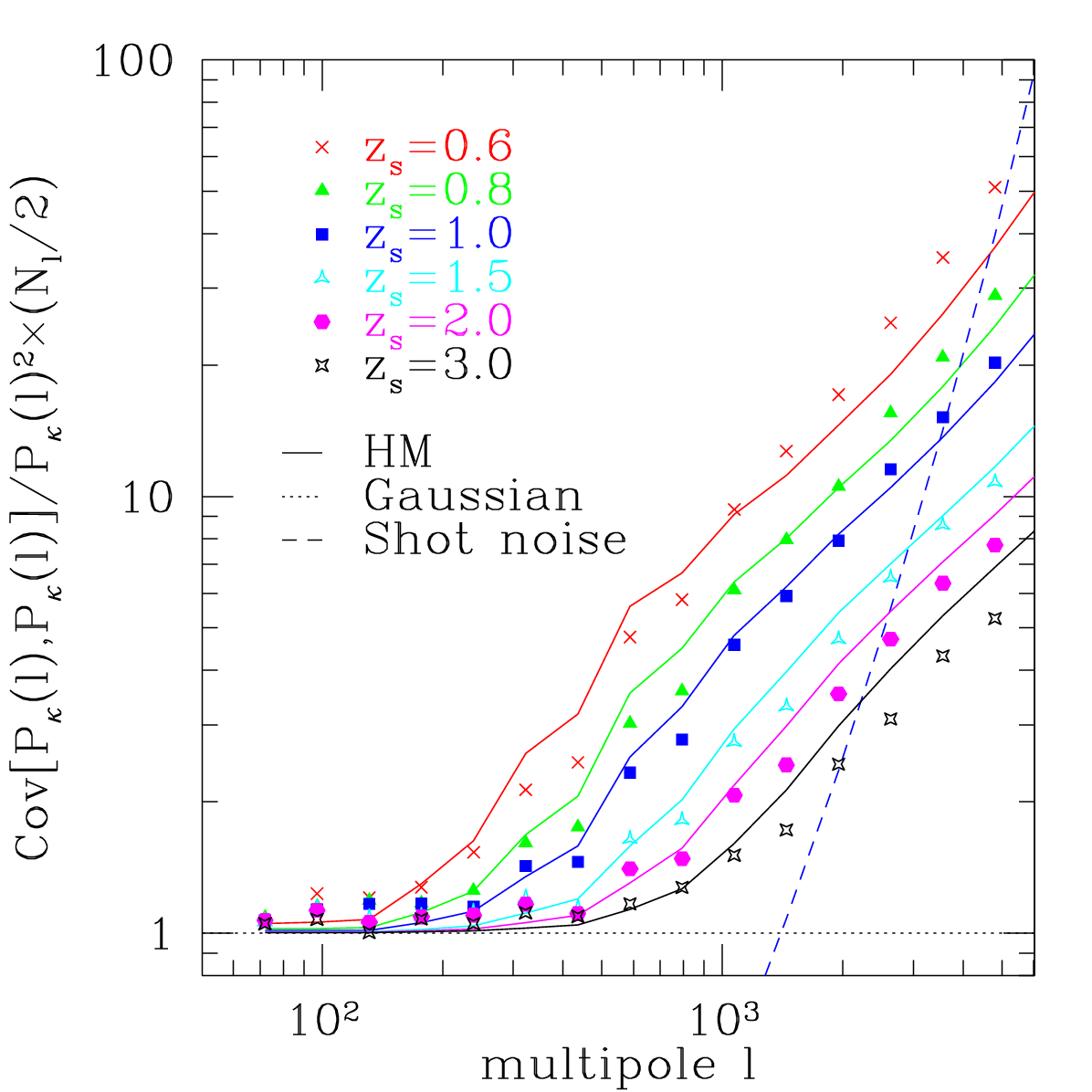}
\caption{The diagonal components of the convergence power spectrum
 covariance for $z_s=0.6,0.8,1.0,1.5,2.0$  and $3.0$, respectively.
 The results are divided by the expected Gaussian covariance (the first term in
 Eq.~(\ref{covariance})).
 Therefore, the deviations from unity arise from the non-Gaussian errors.
 The symbols are the simulation results, while
 the solid curves are the halo model predictions.
The shot noise contribution for
 source redshift $z_s=1.0$ assuming $\bar{n}_g=30$ arcmin$^{-2}$ and 
$\sigma_{\epsilon}=0.22$ for the mean number density and the rms
 intrinsic ellipticities, respectively.}
\label{dia_cov}
\end{figure}

\begin{figure*}
\epsscale{0.95}
\plotone{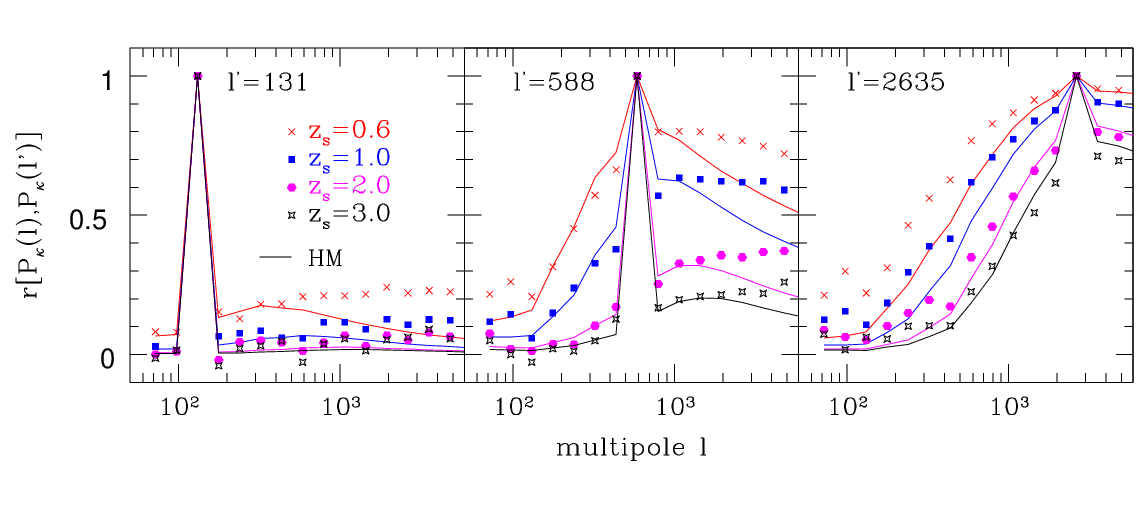}
\caption{The correlation coefficients
 $r[P_{\kappa}(l),P_{\kappa}(l')]$ as a function of $l$ for a given
 $l'$, where $l'$ is chosen to 
$l'=131$
 (left panel), 588 (middle) and $l'=2635$ (right), respectively.
The solid curves denote the halo model predictions. Although the
 simulation and halo model results are in fairly good agreement, 
the simulations results display slightly greater correlation strengths
 for high multipoles and at lower redshifts. 
}
\label{off_cov}
\end{figure*}

\begin{figure*}
\epsscale{0.95}
\plotone{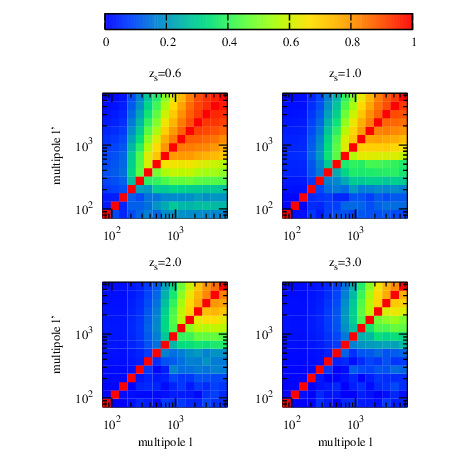}
\caption{Comparison of 
the covariance correlation matrices
predicted by the halo model (upper triangular parts of the
 matrices) and those obtained from
 our 1000 ray-tracing simulations (lower triangular parts)
 for $z_s=0.6,1.0,2.0$ and $3.0$.
}
\label{2dim}
\end{figure*}

\subsection{Diagonal Components of the Covariance Matrix}
Fig.~\ref{dia_cov} shows the diagonal components of the convergence power
 spectrum covariance as a function of multipole.
The values are divided by 
the expected Gaussian covariances for the power spectrum, which is
estimated by inserting the power spectrum measured from the simulations
into the first term on the r.h.s. of Eq.~(\ref{covariance}). 
Therefore, the deviations from unity arise 
from the non-Gaussian error contribution.
The different symbols are the simulation results for different
redshifts. It is clear that the non-Gaussian errors start to be
significant at multipoles from a few hundreds,
 and the
non-Gaussian errors are greater for lower source redshifts due to
stronger nonlinearities in the large-scale structure.

For comparison, the solid curves show the halo model predictions
including the sample variance contribution due to the number
fluctuations of halos in the simulation volume as described around
Eq.~(\ref{eqn:1hsv}). Note that, to obtain the halo model prediction,  
the survey area is set to $\Omega_{\rm
s}=25$ deg$^2$ as assumed for the ray-tracing simulations.  Rather
unexpectedly the empirical 
halo model fairly well reproduces the simulation results
over a wide range of multipoles and for redshifts we have
considered. It should be noted that 
the agreement cannot be found if the sample variance
(\ref{eqn:1hsv}) is not included: the sample variance is dominant over
other non-Gaussian covariance terms at multipoles $l\simgt 1000$ (see
Fig.~\ref{fig:halo_trisp}). 
The $l$-dependence of the
ratio at these high multipoles
is approximately given as ${\rm Cov}^{\rm NG}/{\rm Cov}^{\rm
G}\propto l^1$ for all source redshifts, shallower than the asymptotic
behavior $\propto l^2$ discussed below Eq.~(\ref{eqn:1hsv_asymp}) due to
the residual contributions of other non-Gaussian covariance terms. 

In practice, the shot noise contamination due to intrinsic
galaxy ellipticities contributes to the diagonal term of the
covariance. The dashed line denotes the contribution for source redshift
$z_s=1.0$ assuming $\bar{n}_g=30$ arcmin$^{-2}$ and
$\sigma_{\epsilon}=0.22$ for the mean number density and the rms
intrinsic ellipticities, respectively, which are typical numbers for a
ground-based weak lensing survey such as the planned 
Subaru weak lensing survey. 
It is found that the shot noise becomes significant compared to 
the cosmological
non-Gaussian contributions at very high multipoles, thereby making the
covariance be closer to Gaussian in the multipole range.

\subsection{Off-Diagonal Components of the Covariance Matrix}
The correlation coefficients of the convergence power spectrum
 covariances quantify the relative strengths of the off-diagonal
 components to the diagonal components.
The correlation coefficient is defined as
\begin{equation}
 r[P_{\kappa}(l),P_{\kappa}(l')]=\frac{{\rm Cov}[P_{\kappa}(l),P_{\kappa}(l')]}{\sqrt{{\rm Cov}[P_{\kappa}(l),P_{\kappa}(l)]{\rm
  Cov}[P_{\kappa}(l'),P_{\kappa}(l')]}}.
\end{equation}
Thus the the correlation coefficient is defined so as to give unity when 
$l=l'$. 
For off-diagonal components $r\rightarrow 1$ implies strong correlation
between the two spectra of different multipoles, while $r\rightarrow 0$
means no correlation.
 
Fig.~\ref{off_cov} shows 
the correlation coefficients $r(l,l')$
 as a function of $l$ for a given $l'$ and 
at $z_s=0.6,1.0,2.0$ and $3.0$.
Note that the results depend on the bin width. 
The solid curves denote the halo model predictions.
The halo model predictions fairly well reproduce the simulation results.
A closer look implies a sizable disagreement for very high multipoles
and at lower redshifts. 

Fig.~\ref{2dim} shows 
the correlation coefficient matrices
 at $z_s=0.6,1.0,2.0$ and $3.0$ in a two-dimensional multipole space of 
$(l,l')$.
The upper triangular parts of the matrices are the halo model
 prediction,
while
 the lower triangular parts are the simulation results from our 1000
 realizations.
The correlations are generally stronger at higher $l$ and at lower redshift,
as expected.

\subsection{Signal-to-Noise Ratio}
The obtained covariance matrices can be used to 
estimate the expected signal-to-noise
 $(S/N)$ ratio for measuring the lensing power spectrum.
The cumulative signal-to-noise ratio 
can be defined 
\citep[e.g.][]{2004MNRAS.348..897T,2009MNRAS.395.2065T} as
\begin{equation}
 \left(\frac{S}{N}\right)^2=\sum_{l,l'\le l_{\rm
  max}}P_{\kappa}(l){\rm Cov^{-1}}[P_{\kappa}(l),P_{\kappa}(l')]P_{\kappa}(l'),
\label{signal-to-noise}
\end{equation}
where ${\rm Cov^{-1}}$ is the inverse of the covariance matrix and 
the power spectrum information over $72\simlt l\le l_{\rm max}$ is
included ($l=72$ is the fundamental mode of our ray-tracing simulations,
$l_{\rm f}\simeq 2\pi/5^\circ\simeq 72$).
The signal-to-noise ratio is independent of the
 bin width, as long as the convergence power spectrum does not rapidly
 vary within bin width. 

\begin{figure}
\epsscale{1.0}
\plotone{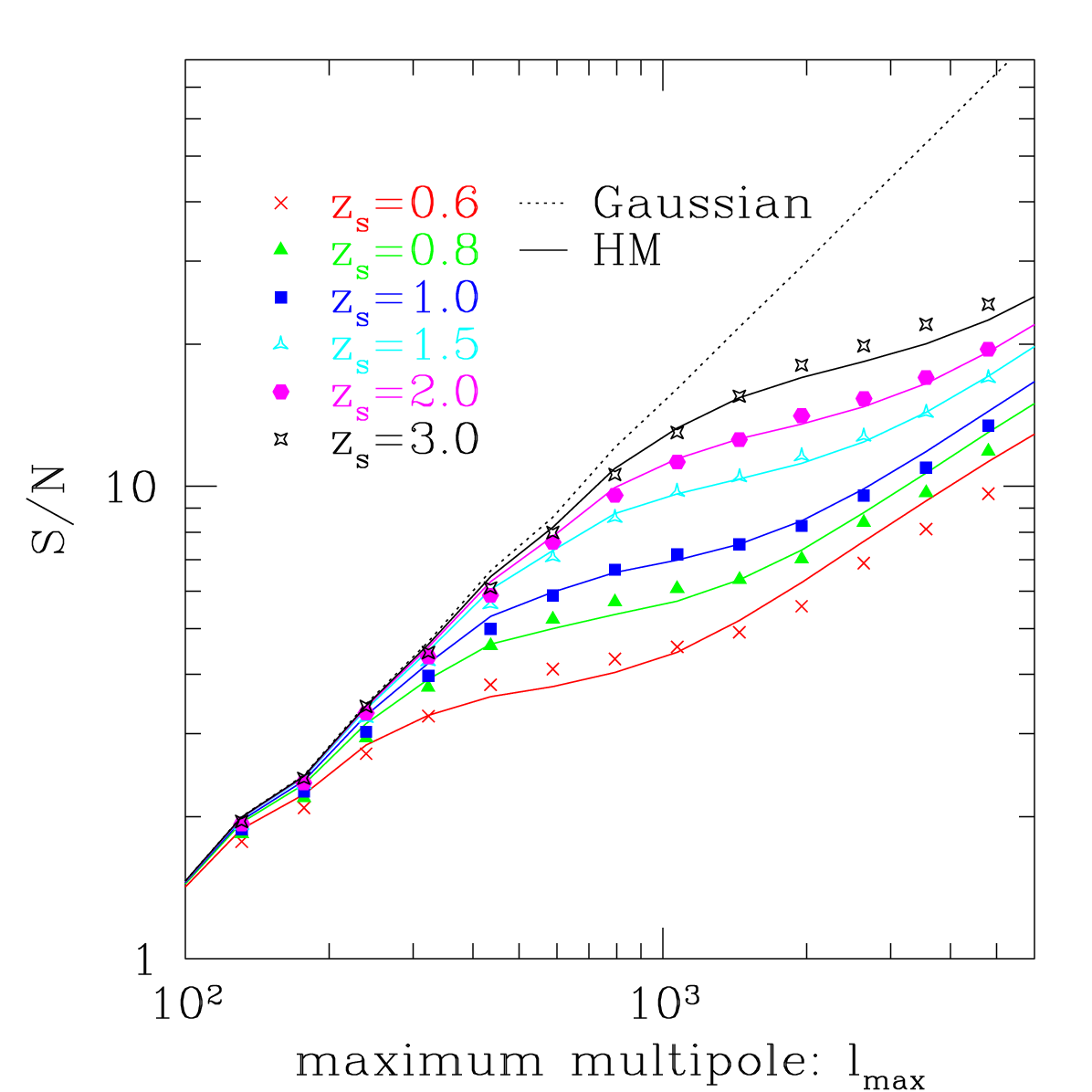}
\caption{The 
cumulative 
signal-to-noise ratio (S/N) 
for the convergence power
 spectrum is shown as a function of maximum multipole $l_{\rm max}$ at
 $z_s=0.6,0.8,1.0,1.5,2.0$ and $3.0$,
where the power spectrum information over a range of multipoles 
$72\le l\le l_{\rm max}$ is included.
The solid curves show the halo model predictions, while the dotted line
 is the result for the Gaussian covariance case. 
}
\label{S_N}
\end{figure}

Fig.~\ref{S_N} shows the $S/N$ for the convergence power spectrum as a
 function of maximum multipole $l_{\rm max}$ for
 $z_s=0.6,0.8,1.0,1.5,2.0$ and $3.0$.
The dotted line shows $S/N$ 
when only Gaussian errors are included.
Our simulation results suggest that the $S/N$ begins to deviate
 significantly from that of the Gaussian case. 
It increases slowly with increasing $l_{\rm max}$ in 
the non-linear regime.
The $S/N$ for low redshift surveys does not increase significantly 
at multipoles from a few hundreds to $1000$. This implies 
that there is little gain in the $S/N$ even if including modes at the
 larger $l$, as has been found in the previous works
 \citep[][]{2005MNRAS.360L..82R,2006MNRAS.371.1188H,2006MNRAS.370L..66N,2007MNRAS.375L..51N,2009ApJ...700..479T,2009arXiv0905.0501D}.
The simulation results show that 
the $S/N$
is degraded by non-Gaussian covariances by up to factor
 5 for source redshift $z_s=1.0$.
\cite{2008ApJ...686L...1L} measured the Fisher information content
for the angular power spectrum of SDSS galaxies,
which is equivalent to the $S/N$ defined in Eq.~(\ref{signal-to-noise}).
They found a similarly significant saturation of the $S/N$. 

The simulation results can be compared with the halo model predictions,
where the non-Gaussian errors are computed from the halo model, and 
 the power spectrum and the Gaussian
term of covariance are taken from the simulation results in the $S/N$
calculation. 
The halo model predictions are again in a good agreement 
 with the simulation results, and in particular 
well capture complex
 dependences of the $S/N$ on multipoles and source redshifts. 

For planned weak lensing surveys most important range of multipoles are
around $l\sim 1000$ in order to avoid complications due to effects of strong
nonlinear clustering and gas dynamics on mass power spectrum. Over such a
range of multipoles the cosmological non-Gaussian errors 
are dominant over the shot
noise due to intrinsic galaxy shapes, therefore the effect needs to be
properly taken into account to obtain unbiased, secure extraction of
cosmological parameters from the measured power spectrum (see \cite{2009MNRAS.395.2065T} for the similar discussion). 

\section{Probability Distribution of the Convergence Power Spectrum
 Estimator}
\label{sec:prob}

\begin{figure*}
\epsscale{1.0}
\plottwo{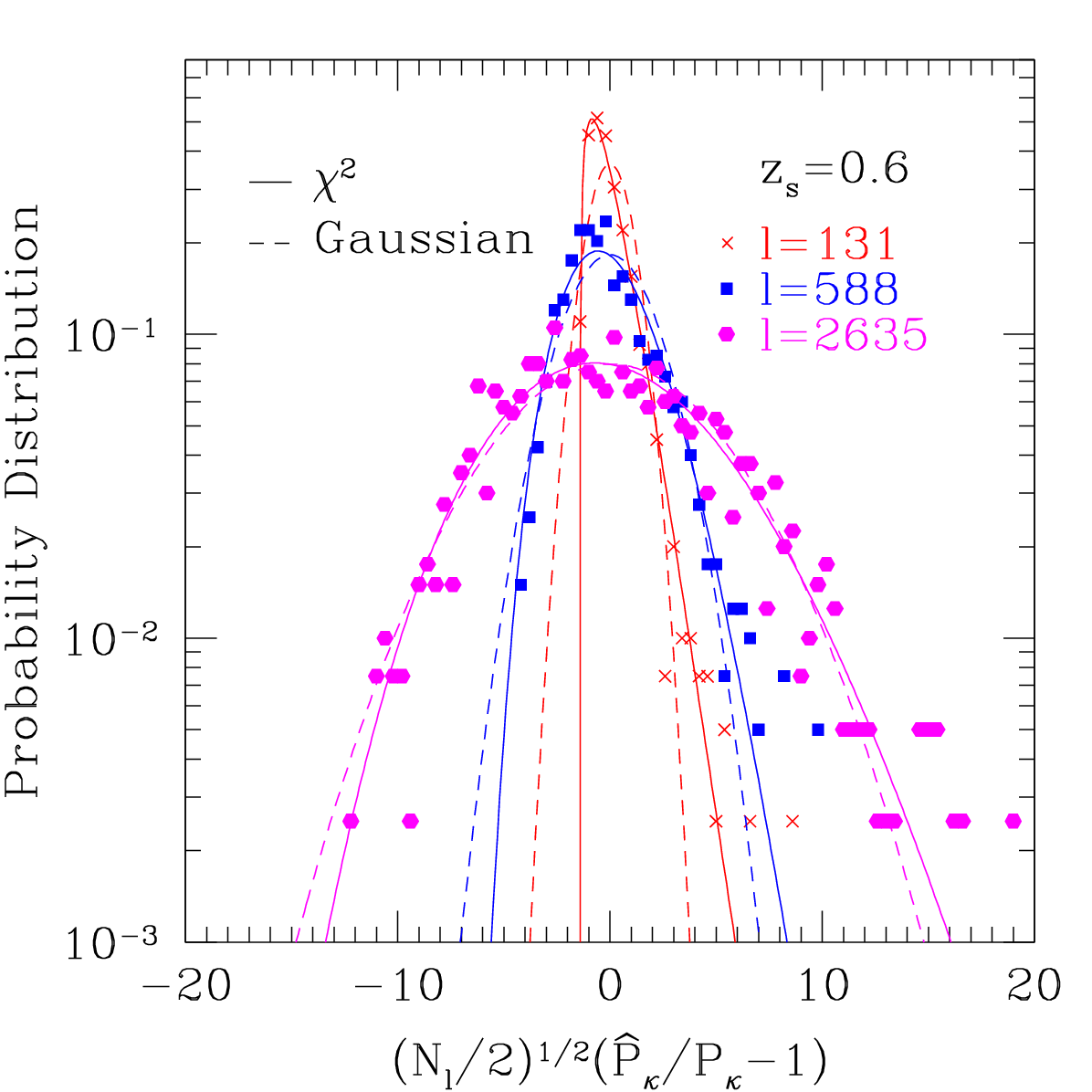}{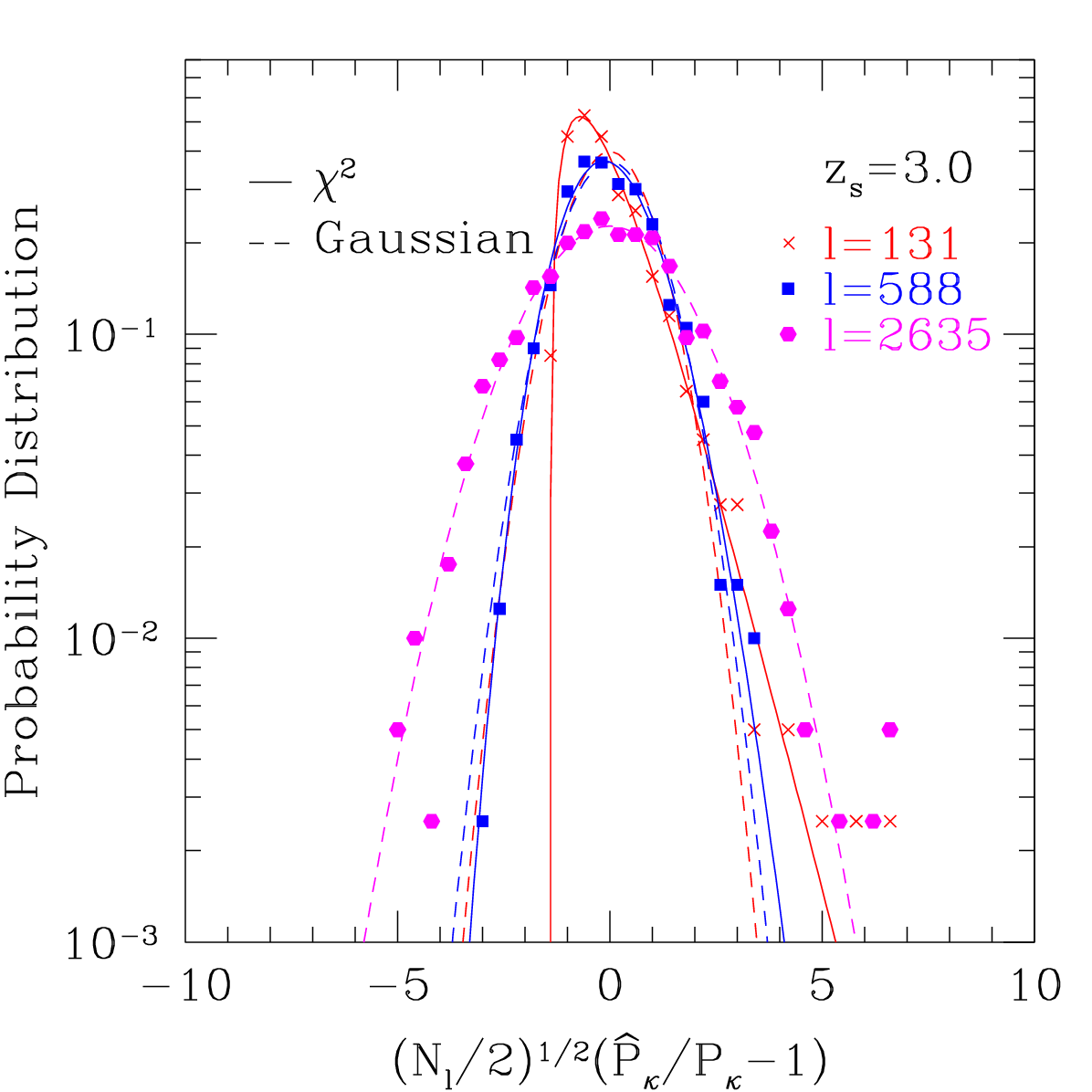}
\caption{Probability distribution of the convergence power spectrum
 estimators $\hat{P}_{\kappa}$ among the 1000 realizations for
 $z_s=0.6$ (left panel) and  $3.0$ (right), respectively. The solid and dashed 
 curves show the $\chi^2$- and Gaussian-distributions with zero mean, 
respectively,
 where the variance is set to the 
to the diagonal covariance components measured from the
 simulations.
}
\label{proba}
\end{figure*}

In this section, we study the probability distribution of the
 convergence power spectrum estimator $\hat{P}_{\kappa}$
in order to see how the non-linear growth causes a
 non-Gaussian distribution in the convergence power spectrum estimators.
It should be noted that the convergence power spectrum covariance simply
 reflects the width of the full distribution at each $l$.

Fig.~\ref{proba} shows the probability distribution of the convergence
 power spectrum estimators $\hat{P}_{\kappa}$ among 1000 realizations.
We measure the probability distribution for $(N_l/2)^{1/2}(\hat{P}_{\kappa}/P_{\kappa}-1)$
so that the mean and variance of the distribution are equals to
 zero and unity in the linear regime 
or if the convergence field is
 Gaussian.
The probability distribution is normalized 
so as to give unity when integrated over all $\hat{P}_\kappa$ values. 
For large $l$ values, the distribution is broadened due to non-linear evolution.
The solid and dashed curves show the $\chi^2$-distribution and 
the expected Gaussian
 distribution, respectively, where the variance for each of the
 distributions is set to the 
diagonal term of covariance measured from simulations at
 each $l$, i.e. the variance includes the non-Gaussian covariance
 contribution as given in Fig.~\ref{dia_cov}. To be more explicit, 
assuming that the estimate $\hat{P}_{\kappa}$ obeys the $\chi^2$-distribution,
the mean and variance are set to $P_{\kappa}(l)$ and
 Cov[$P_{\kappa}(l),P_{\kappa}(l)$] (replacing
 $\hat{P}(k)\rightarrow \hat{P}_{\kappa}(l)$, 
$N_k/2\rightarrow P_{\kappa}^2(l)/{\rm
 Cov}[P_{\kappa}(l),P_{\kappa}(l)]$ in Eq.~(B1) in
 \cite{2009ApJ...700..479T}).
Fig.~\ref{proba} shows that 
the probability distribution is well approximated by
the $\chi^2$-distribution, but display a larger positive tail than
expected from the $\chi^2$-distribution.  
One can see that the estimators have a skewed distribution, 
even for the low multipoles such as $l\sim 100$ where the lensing fields
are more in the linear regime.

The deviations from a Gaussian distribution 
can be quantified by studying 
 skewness $S_3$ and kurtosis $S_4$ 
of the distribution in Fig.~\ref{proba}: 
 \begin{align}
&S_3=\frac{\langle(\hat{P}_{\kappa}(l)-P_{\kappa}(l))^3\rangle}{\langle(\hat{P}_{\kappa}(l)-P_{\kappa}(l))^2\rangle^{3/2}}, \nonumber\\
&S_4=\frac{\langle(\hat{P}_{\kappa}(l)-P_{\kappa}(l))^4\rangle}{\langle(\hat{P}_{\kappa}(l)-P_{\kappa}(l))^2\rangle^{2}}-3 .
\end{align}
If the convergence field is a Gaussian random field, which is a
 good approximation in linear regime, the convergence power spectrum
 estimator $\hat{P}_{\kappa}$ of a given $l$ exactly obeys the
 $\chi^2$-distribution.
In this case, 
the skewness and
 kurtosis can be analytically computed as
\begin{equation}
 S_3=\sqrt{\frac{4\, {\rm
  Cov}[P_{\kappa}(l),P_{\kappa}(l)]}{P_{\kappa}(l)^2}},\quad S_4=\frac{6\,
  {\rm Cov}[P_{\kappa}(l),P_{\kappa}(l)]}{P_{\kappa}(l)^2}.
\label{theo_sk_ku}
\end{equation}
Note that $S_3$ and $S_4$ scale with survey area as 
$S_3\propto \Omega_{\rm s}^{-1/2}$ and $S_4\propto \Omega_{\rm s}^{-1}$,
as ${\rm Cov}\propto \Omega_{\rm s}^{-1}$. 

\begin{figure}
\epsscale{1.0}
\plotone{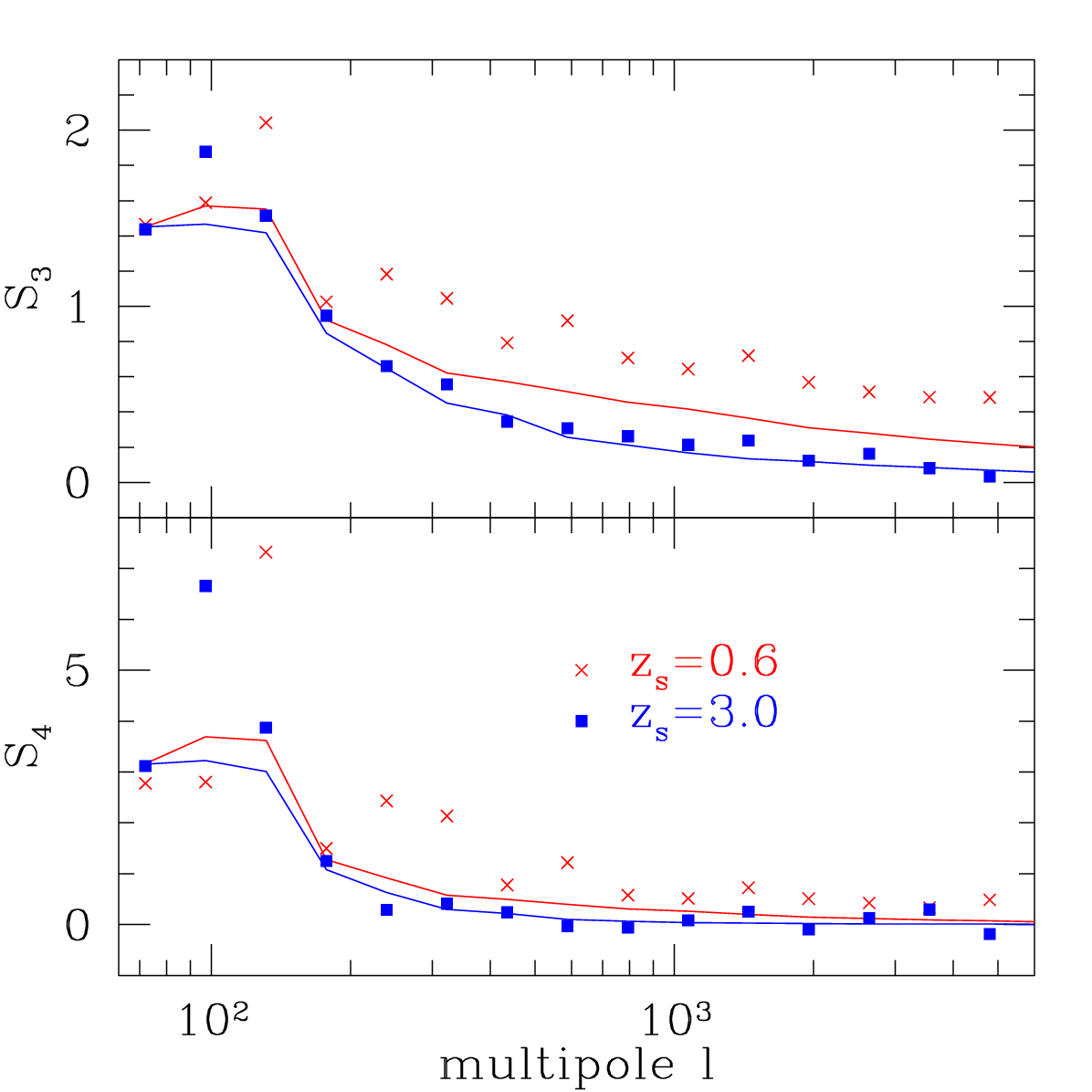}
\caption{The skewness (top panel) and the kurtosis (bottom) of the
 convergence power spectrum distribution as a function of multipole $l$
 at $z_s$=0.6 and 3.0. The solid curves are the theoretical predictions
from Eq.~(\ref{theo_sk_ku}).} 
\label{sk_ku}
\end{figure}

Fig.~\ref{sk_ku} shows the simulation results for 
$S_3$ and $S_4$ as a function
 of multipole $l$.
Note that the results in Fig.~\ref{sk_ku} are for a survey area of 
$\Omega_{\rm s}=25$ degree$^2$.
The solid curves are the theoretical predictions of
Eq.~(\ref{theo_sk_ku}). 
The model well describes the simulation results for $z_s=3.0$, while the
results for $z_s=0.6$ show significant non-Gaussian cumulants over a
range of multipoles we have considered, 
due to stronger non-linearities. 
For $z_s=3.0$, both $S_3$ and $S_4$ asymptote to zero at high $l$,
 i.e. the probability distribution approaches to a Gaussian distribution
 at high $l$ due to the central limit theorem.

Since the skewness and the kurtosis have non-negligible values 
at multipoles relevant for future surveys, 
a prior knowledge on the full distribution
may be needed to obtain an unbiased estimate on the ensemble averaged
band power at each $l$ bin. 

\section{Conclusion and Discussion}
\label{sec:conc}
Accurate statistics are essential in 
the likelihood analysis for future precision cosmology.
We can exploit the full potential of upcoming high quality data, only 
if we use appropriate statistical methods.
For weak-lensing surveys, non-linear gravitational 
evolution of large-scale structure can significantly compromise 
 cosmological parameter estimations and thus
 needs to be modelled accurately.

We have used ray-tracing simulations
in order to study how the non-Gaussian covariance varies
 with scales and redshifts for the
 standard $\Lambda$CDM cosmology.
We have performed a total of 1000 independent ray-tracing simulations using 400
 cosmological $N$-body simulations.
The non-Gaussian errors become more
 significant on smaller scales and at lower redshifts.
The cumulative signal-to-noise ratio ($S/N$) for measuring lensing power
 spectrum is degraded due to non-Gaussian covariance by up to a factor of 
5 for a weak lensing survey to $z_s\sim 1$.
We show that the simulation results are fairly well described by 
the halo model prediction including 
additional contribution due to the
 statistical fluctuations in the number of halos in a finite
survey volume.

We also study the probability distribution of the convergence power
 spectrum estimator among 1000 realizations. 
The probability distribution has a large skewness especially for shallow
 surveys, which is likely due to nonlinear gravitational evolution.
Therefore, a prior knowledge on the full distribution may be needed to
 obtain an unbiased estimate on the ensemble averaged band power at each $l$.
Overall, the non-Gaussian errors likely cause best-fitting
 parameters to be systematically biased, 
if the model fitting is done improperly assuming the Gaussian covariances.
Therefore it is clearly needed to develop 
an appropriate method for parameter estimations from the
measured power spectrum taking into account 
the non-Gaussian errors. 

The most conventionally used statistical measure is the
cosmic shear correlation function.
An invaluable feature of the correlation function is that it does not
 require non-trivial corrections for survey geometry and masking effects.
Therefore, it is useful to estimate the covariance matrix of real space
 correlation function and to derive fitting formula to calibrate the
 full covariances for an arbitrary survey area.
These issues will be studied in a subsequent paper.
  
\acknowledgments
We would like to thank Issha Kayo for useful comments and discussions.
We also thank the anonymous referee for careful reading of our
manuscript and very useful suggestions.
This work is supported in part by 
by World Premier International Research Center
Initiative (WPI Initiative), and by 
Grant-in-Aid for Scientific Research
 on Priority Areas No. 467 ``Probing the Dark Energy through an
 Extremely Wide and Deep Survey with Subaru Telescope'' and by the
 Grant-in-Aid for Nagoya University Global COE Program, ``Quest for
 Fundamental Principles in the Universe: from Particles to the Solar
 System and the Cosmos'', from the Ministry of Education, Culture,
 Sports, Science and Technology of Japan.
Numerical computations were in part carried out on the general-purpose
PC farm at Center for Computational Astrophysics, CfCA, of National
Astronomical Observatory of Japan.

\bibliography{ms}

\clearpage

\end{document}